\documentclass[12pt]{article}
\usepackage{amstex}

\def\Htilde{\tilde{H}}
\def\Ytilde{\tilde{Y}}

\def\ep{\epsilon}

\begin{document}
\vspace*{-.6in}
\thispagestyle{empty}
\begin{flushright}
CALT-68-2115\\
hep-th/9705092
\end{flushright}
\baselineskip = 20pt

\vspace{.5in}
{\Large
\begin{center}
The M Theory Five-Brane and the Heterotic String\footnote{Talk presented at the
{\it Trieste Conference on Duality Symmetries in String Theory - II}}
\end{center}}

\begin{center}
John H. Schwarz\\
\emph{California Institute of Technology, Pasadena, CA  91125, USA}
\end{center}
\vspace{0.1in}

\begin{center}
\textbf{Abstract}
\end{center}
\begin{quotation}
\noindent  Brane actions with chiral bosons present special challenges. Recent
progress in the description of the two main examples -- the M theory five-brane
and the heterotic string -- is described. Also, double dimensional reduction of the
M theory five-brane on K3 is shown to give the heterotic string.
\end{quotation}
\bigskip

\section{The Bosonic Part of the Five-Brane Action}

The M theory five-brane action contains a tensor gauge field, which (in linearized 
approximation) has a self-dual field strength.
Ref. \cite{jhs} analyzed the problem of coupling a 6d self-dual tensor gauge field to a
metric field so as to achieve general coordinate invariance. 
It presented a formulation in which one direction is treated differently from
the other five. At the time that work was done,
the author knew of no straightforward way to make the general
covariance manifest. However, shortly thereafter a paper appeared~\cite{pasti1} that 
presents equivalent results using a manifestly covariant formulation~\cite{pasti2},
which we refer to as the PST formulation. In the following both approaches and their
relationship are described. These results have been generalized to
supersymmetric actions with local kappa symmetry~\cite{bandos,aganagic,howe}, 
but here we will only consider the bosonic theories.

\medskip

\noindent{\it The Noncovariant Formulation}

Let us denote the 6d (world volume) coordinates by 
$\sigma^{\hat\mu} = (\sigma^\mu, \sigma^5)$,
where $\mu = 0,1,2,3,4$. The $\sigma^5$ direction is singled out as the one that
will be treated differently from the other five.\footnote{This is a
space-like direction, but one could also choose a time-like  one.}  The 6d metric
$G_{\hat\mu\hat\nu}$ contains 5d pieces $G_{\mu\nu}, G_{\mu 5}$, and $G_{55}$.
All formulas will be written with manifest 5d general coordinate invariance.
As in refs.~\cite{perry,jhs}, we represent the self-dual tensor gauge field by a
$5\times 5$ antisymmetric tensor $B_{\mu\nu}$, and its 5d curl by
$H_{\mu\nu\rho} = 3 \partial_{[\mu} B_{\nu\rho]}$. A useful quantity is the dual 
\begin{equation}
\tilde{H}^{\mu\nu} = {1\over 6} \epsilon^{\mu\nu\rho\lambda\sigma}
H_{\rho\lambda\sigma}.
\end{equation}

It was shown in ref.~\cite{jhs} that a class of generally covariant
bosonic theories can be represented in the form
$L = L_1 + L_2 + L_3$, where
\begin{eqnarray}
L_1 &=& -{1\over 2}\sqrt{-G} f(z_1,z_2), \nonumber \\
L_2 &=& -{1\over 4} \tilde{H}^{\mu\nu} \partial_5 B_{\mu\nu}, \\
L_3 &=&  {1\over 8}
\epsilon_{\mu\nu\rho\lambda\sigma} {G^{5\rho}\over G^{55}} \tilde{H}^{\mu\nu}
\tilde{H}^{\lambda\sigma}.\nonumber 
\end{eqnarray}
The notation is as follows:  $G$ is the 6d determinant $(G =
{\rm det}\, G_{\hat\mu\hat\nu})$ and
$G_5$ is the 5d determinant $(G_5 =
{\rm det}\, G_{\mu\nu})$, while $G^{55}$ and $G^{5\rho}$ are components of the inverse
6d metric $G^{\hat\mu \hat\nu}$.  The $\epsilon$ symbols are purely numerical with $\epsilon^{01234} = 1$ and
$\epsilon^{\mu\nu\rho\lambda\sigma} = - \epsilon_{\mu\nu\rho\lambda\sigma}$.  A
useful relation is $G_5 = G G^{55}$.
The $z$ variables are defined to be
\begin{eqnarray}
z_1 &=& {{\rm tr} (G\tilde{H} G\tilde{H})\over 2( -G_5)}\nonumber \\
z_2 &=& {{\rm tr} (G\tilde{H} G\tilde{H} G\tilde{H} G\tilde{H})\over 4 (-G_5)^2}.
\label{zdefs}
\end{eqnarray}
The trace only involves 5d indices:
\begin{equation}
{\rm tr} (G\tilde{H} G\tilde{H}) = G_{\mu\nu} \tilde{H}^{\nu\rho} G_{\rho\lambda}
\tilde{H}^{\lambda\mu}.
\end{equation}
The quantities $z_1$ and $z_2$
are scalars under 5d general coordinate
transformations.  

Infinitesimal parameters of general coordinate transformations are denoted
$\xi^{\hat\mu} = (\xi^\mu, \xi)$.  Since 5d general coordinate invariance is
manifest, we focus on the $\xi$ transformations only.  The metric transforms in
the standard way
\begin{equation}
\delta_\xi G_{\hat\mu \hat\nu} = \xi \partial_5 G_{\hat\mu \hat\nu} +
\partial_{\hat\mu} \xi G_{5\hat\nu} + \partial_{\hat\nu} \xi G_{\hat\mu 5}.
\label{Gvar}
\end{equation}
The variation of $B_{\mu\nu}$ is given by a more complicated rule, whose origin is
explained in ref.~\cite{jhs}:
\begin{equation}
\delta_\xi B_{\mu\nu} = \xi K_{\mu\nu}, \label{Bvar}
\end{equation}
where
\begin{equation}
K_{\mu\nu} = 2{\partial (L_1 + L_3) \over\partial \tilde{H}^{\mu\nu}} =
K_{\mu\nu}^{(1)} f_1+ K_{\mu\nu}^{(2)} f_2+ K_{\mu\nu}^{(\epsilon)}
\label{Kform1}
\end{equation}
with
\begin{eqnarray}
K_{\mu\nu}^{(1)} &=& {\sqrt{-G} \over (-G_5)}{(G\tilde{H} G)_{\mu\nu}}
\nonumber \\
K_{\mu\nu}^{(2)} &=& {\sqrt{-G} \over (-G_5)^2}{(G\tilde{H} G\tilde{H} G\tilde{H}
G)_{\mu\nu}}  \label{Kform2}\\
K_{\mu\nu}^{(\epsilon)} &=& \epsilon_{\mu\nu\rho\lambda\sigma}
{G^{5\rho}\over 2 G^{55}} \tilde{H}^{\lambda\sigma}, \nonumber
\end{eqnarray}
and we have defined
\begin{equation}
f_i = {\partial f\over\partial z_i} , \quad i = 1,2.
\end{equation}

Assembling the results given above, ref.~\cite{jhs} showed that
the required general coordinate transformation symmetry is
achieved if, and only if, the function $f$ satisfies the nonlinear partial
differential equation~\cite{gibbons}
\begin{equation}
f_1^2 + z_1 f_1 f_2 + \big({1\over 2} z_1^2 - z_2\big) f_2^2 = 1.
\end{equation}
As discussed in~\cite{perry},
this equation has many solutions, but the one of relevance to the
M theory five-brane is 
\begin{equation}
f = 2 \sqrt{1 + z_1 + {1\over 2} z_1^2 - z_2}.
\end{equation}
For this choice $L_1$
can reexpressed in the Born--Infeld form
\begin{equation}
L _1 = - \sqrt{- {\rm det} \Big(G_{\hat\mu \hat\nu} + i G_{\hat\mu\rho} G_{\hat\nu
\lambda} \tilde{H}^{\rho\lambda} / \sqrt{-G_5}\Big)} . \label{bosonicL1}
\end{equation}
This expression is real, despite the factor of $i$, because it is an even function of
$\tilde H$.

\medskip

\noindent{\it The PST Formulation}

In ref.~\cite{pasti1} (using techniques developed
in ref.~\cite{pasti2}) equivalent results are
described in a manifestly covariant way.  To do this, the field $B_{\mu\nu}$ is
extended to $B_{\hat\mu \hat\nu}$ with field strength $H_{\hat\mu \hat\nu
\hat\rho}$.  In addition, an auxiliary scalar field $a$ is
introduced.  The PST formulation has new gauge symmetries (described below)
that allow one to choose the gauge $B_{\mu 5} = 0,$  $a = \sigma^5$
(and hence $\partial_{\hat\mu}a =
\delta_{\hat\mu}^5$).  In this gauge, the covariant PST formulas reduce to the
ones given above.

Equation (\ref{bosonicL1}) 
expresses $L_1$ in terms of the determinant of the $6 \times 6$ matrix
\begin{equation}
M_{\hat\mu\hat\nu} = G_{\hat\mu\hat\nu} + i {G_{\hat\mu \rho} G_{\hat\nu
\lambda}\over \sqrt{- GG^{55}}} \tilde{H}^{\rho\lambda}.
\end{equation}
In the PST approach this is extended to the manifestly covariant form
\begin{equation}
M_{\hat\mu\hat\nu}^{\rm cov.} = G_{\hat\mu\hat\nu} + i {G_{\hat\mu\hat\rho}
G_{\hat\nu \hat\lambda}\over\sqrt{-G (\partial a)^2}}
\tilde{H}_{\rm cov.}^{\hat\rho \hat\lambda}. \label{Mcov}
\end{equation}
The quantity
\begin{equation}
(\partial a)^2 = G^{\hat\mu\hat\nu} \partial_{\hat\mu} a \partial_{\hat\nu} a
\end{equation}
reduces to $G^{55}$ upon setting $\partial_{\hat\mu}a  = \delta_{\hat\mu}^5$,
and
\begin{equation}
\tilde{H}_{\rm cov.}^{\hat\rho\hat\lambda} \equiv {1\over 6} \epsilon^{\hat\rho
\hat\lambda \hat\mu \hat\nu \hat\sigma \hat\tau} H_{\hat\mu \hat\nu \hat\sigma}
\partial_{\hat\tau} a
\end{equation}
reduces to $\tilde{H}^{\rho\lambda}$.  Thus $M_{\hat\mu \hat\nu}^{\rm cov.}$
replaces $M_{\hat\mu\hat\nu}$ in $L_1$.  Furthermore, the expression
\begin{equation}
L' = - { 1\over 4(\partial a)^2} \tilde{H}_{\rm cov.}^{\hat\mu \hat\nu}
H_{\hat\mu\hat\nu\hat\rho} G^{\hat\rho\hat\lambda} \partial_{\hat\lambda} a,
\end{equation}
which transforms under general coordinate transformations as a scalar density,
reduces to $L_2 + L_3$ upon gauge fixing. It is interesting that $L_2$ and $L_3$ are
unified in this formulation.

Let us now describe the new gauge symmetries of ref.~\cite{pasti1}.  Since degrees of
freedom $a$ and $B_{\mu 5}$ have been added, corresponding gauge symmetries are
required.  One of them is
\begin{equation}
\delta B_{\hat\mu \hat\nu} = 2 \phi_{[\hat\mu} \partial_{\hat\nu]} a,
\end{equation}
where $\phi_{\hat\mu}$ are infinitesimal parameters, and the other fields do not
vary.  In terms of differential forms, this implies $\delta H = d\phi\wedge
da$.  $\tilde{H}_{\rm cov.}^{\hat\rho \hat\lambda}$ is invariant under this transformation,
since it corresponds to the dual of $H\wedge da$, but $da\wedge da = 0$.
Thus the covariant version of $L_1$ is invariant under this transformation.
The variation of $L'$, on the other hand, is a total derivative.

The second local symmetry involves an infinitesimal 
scalar parameter $\varphi$.  The transformation
rules are $\delta G_{\hat\mu\hat\nu} = 0, \delta a = \varphi$, and
\begin{equation}
\delta B_{\hat\mu\hat\nu} = {1\over (\partial a)^2} \varphi
H_{\hat\mu\hat\nu\hat\rho} G^{\hat\rho\hat\lambda} \partial_{\hat\lambda} a +
\varphi V_{\hat\mu\hat\nu},
\end{equation}
where the quantity $V_{\hat\mu\hat\nu}$ is to be determined.   
Rather than derive it from
scratch, let's see what is required to agree with the previous formulas after
gauge fixing.  In other words, we fix the gauge $\partial_{\hat\mu} a =
\delta_{\hat\mu}^5$ and $B_{\mu 5} = 0$, and figure out what the resulting
$\xi$ transformations are.  We need
\begin{equation}
\delta a = \varphi + \xi \partial_5 a = \varphi + \xi = 0,
\end{equation}
which tells us that $\varphi = - \xi$.  Then
\begin{eqnarray}
\delta_{\xi} B_{\mu\nu} &=& {1\over (\partial a)^2} \varphi H_{\mu\nu\hat\rho}
G^{\hat\rho\hat\lambda} \partial_{\hat\lambda} a + \varphi V_{\mu\nu} + \xi
H_{5\mu\nu}\nonumber \\
&=& - \xi \left({G^{\rho 5}\over G^{55}} H_{\mu\nu\rho} + V_{\mu\nu}\right) =
\xi (K_{\mu\nu}^{(\epsilon)} - V_{\mu\nu}).
\end{eqnarray}
Thus, comparing with eqs.~(\ref{Bvar}) and (\ref{Kform1}), we need the covariant definition
\begin{equation}
V_{\hat\mu\hat\nu} = - 2 {\partial L_1\over \partial
\tilde{H}_{\rm cov.}^{\hat\mu\hat\nu}}
\end{equation}
to achieve agreement with our previous results.

\medskip

\section{A New Heterotic String Action}

There are two main approaches to constructing the world-sheet action of the
heterotic string that have been used in the past~\cite{gross}.  
In one of them, the internal
torus is described in terms of bosonic coordinates. The fact that these
bosons are chiral ({\it i.e.}, the left-movers and right-movers behave differently)
is imposed through external constraints.  In the second approach these bosonic
coordinates are replaced by world-sheet fermions, which are Majorana--Weyl in
the 2d sense.  What will be most convenient for our purposes is a variant of
the first approach.  In this variant the coordinates of the Narain torus are still
represented by bosonic fields, but the chirality of these fields is achieved
through new gauge invariances rather than external constraints~\cite{cherkis}.

Consider the Narain compactified heterotic string in a
Minkowski space-time with $d = 10 - n$ dimensions~\cite{narain}.  
Let these coordinates be denoted
by $X^m$ with $m = 0,1, \ldots, d - 1 = 9 - n$.  To properly account for all
the degrees of freedom, the Narain torus should be described by $16 + 2n$
bosonic coordinates $Y^I, $ $I = 1,2, \ldots, 16 + 2n$.  These will be arranged to
describe $26 - d = 16 + n$ left-movers and $10 - d = n$ right-movers.  The
$Y^I$ are taken to be angular coordinates, with period $2\pi$, so that $Y^I
\sim Y^I + 2\pi$, and the conjugate momenta are integers.  The actual size and
shape of the torus is encoded in a matrix of moduli, denoted $M_{IJ}$, which
will be described below.

The $(16 + 2n)$-dimensional
lattice of allowed momenta should form an even self-dual lattice of signature
$(n, 16 + n)$.  Let us therefore introduce a matrix
\begin{equation}
\eta = \left(\begin{matrix}
I_n & 0\\
0 & - I_{16+n}\end{matrix}\right) ,
\end{equation}
where $I_n$ is the $n \times n$ unit matrix.  An
even self-dual lattice with this signature has a set of $16 + 2n$ basis
vectors $V_I$, and the symmetric matrix
\begin{equation}
L_{IJ} = V_I^a \eta_{ab} V_J^b
\end{equation}
characterizes the lattice.  A convenient specific choice is
\begin{equation}
L = \Lambda_8 \oplus \Lambda_8 \oplus \sigma \oplus \ldots \oplus \sigma,
\end{equation}
where $\Lambda_8$ is the negative of the $E_8$ Cartan matrix and $\sigma =
\left(\begin{matrix} 0 & 1\\ 1 & 0 \end{matrix}\right)$ appears $n$ times. 

The Narain moduli space is characterized, up to $T$ duality equivalences that
will be discussed below, by a symmetric matrix $M'_{ab} \in O (n, 16 +
n)$, which satisfies $M'\eta M' = \eta$. 
The fact that it is symmetric means that it actually parametrizes the
coset space $O(n, 16 + n)/O(n) \times O (16+n)$, which has $n(16 + n)$ real
dimensions.  To describe the $T$ duality equivalences, it is convenient to
change to the basis defined by the basis vectors of the self-dual lattice.
Accordingly, we define
\begin{equation}
M_{IJ} = V_I^a M'_{ab} V_J^b = (V^T M' V)_{IJ}.
\end{equation}
This matrix is also symmetric and satisfies
\begin{equation}
ML^{-1} M = L,
\end{equation}
from which it follows that $(L^{-1}M)^2 = 1$.  This allows us to define
projection operators
\begin{equation}
{\cal P}_\pm = {1\over 2} (1 \pm L^{-1} M).
\end{equation}
${\cal P}_+$ projects onto an $n$-dimensional subspace, which will correspond
to right-movers.  Similarly, ${\cal P}_-$ projects onto the
$(16+n)$-dimensional space of left-movers.
The theory we are seeking should be 
invariant under an infinite discrete group of $T$ duality
transformations, denoted 
$\Gamma_{n,16+n}$,\footnote{It is often called $O(n,16 + n; {\bf Z})$.} 
so that the actual moduli space is the standard Narain space
\begin{equation}
{\cal M}_{n, 16 + n} = \Gamma_{n, 16 + n}\backslash O(n, 16 + n) / O(n) \times O (16 +n).
\end{equation}

The desired equations of motion for the $Y$ 
coordinates are~\cite{cecotti,duff,tseytlin,maharana}
\begin{equation}
{\cal P}_- \partial_+ Y = 0 \quad { \rm and} \quad {\cal P}_+ \partial_- Y = 0,
\label{Yeqs}
\end{equation}
where $\xi^\pm = \xi^1 \pm \xi^0$, so that $\partial_\pm = {1\over 2}
(\partial_1 \pm \partial_0)$.  $\xi^0$ and $\xi^1$ are the world-sheet time and
space, respectively. The pair of equations in (\ref{Yeqs}) can be combined
in the form
\begin{equation}
M\partial_0 Y - L\partial_1 Y = 0.
\end{equation}
It is easy to write down a lagrangian that gives this equation~\cite{floreanini}:
\begin{equation}
{\cal L}_N = {1\over 2} (\partial_0 Y M \partial_0 Y - \partial_0 Y L \partial_1 Y).
\end{equation}
Two things are peculiar about this lagrangian.  First, it does not
have manifest Lorentz invariance.  
However, in ref.~\cite{schwarz} it
was shown that ${\cal L}_N$ has a global symmetry that can be interpreted as
describing a non-manifest Lorentz invariance. 
Second, it gives the equation of motion
\begin{equation}
\partial_0 [M\partial_0 Y - L\partial_1 Y] = 0,
\end{equation}
which has a second, unwanted, solution $Y^I = f^I (\xi^1)$.  The resolution of
the second problem is quite simple.  The transformation $\delta Y^I = f^I
(\xi^1)$ is a gauge symmetry of ${\cal L}_N$, and therefore $f^I (\xi^1)$ represents
unphysical gauge degrees of freedom.

The first problem, the noncovariance of ${\cal L}_N$, is more interesting.  
We will follow the PST approach~\cite{pasti2}, 
and extend ${\cal L}_N$ to a manifestly Lorentz
invariant action by introducing an auxiliary scalar field $a(\xi)$.  The
desired generalization of ${\cal L}_N$ is then
\begin{equation}
{\cal L}_{PST} = {1\over 2(\partial a)^2} (\tilde{Y} M \tilde{Y} - \tilde{Y} L\,
\partial Y \cdot \partial a),
\end{equation}
where
\begin{equation}
\tilde{Y}^I = \epsilon^{\alpha\beta} \partial_\alpha Y^I \partial_\beta a.
\label{Ytilde}
\end{equation}
Also, $(\partial a)^2$ and $ \partial Y \cdot \partial a$ are formed using the 2d
Lorentz metric, which  is diagonal with $\eta^{00} = - 1$ and $\eta^{11} = 1$.  

The theory given by ${\cal L}_{PST}$ has two gauge invariances.  The first is
\begin{eqnarray}
\delta Y &=& \varphi \left({1\over \partial_+ a} {\cal P}_- \partial_+ Y +
{1\over\partial_- a} {\cal P}_+ \partial_- Y\right),\nonumber \\
\delta a &=& \varphi,
\end{eqnarray}
where $\varphi (\xi^0, \xi^1)$ is an arbitrary infinitesimal scalar function.
If this gauge freedom is used to set $a = \xi^1$, then ${\cal L}_{PST}$ reduces to
${\cal L}_N$.  The second gauge invariance is
\begin{equation}
\delta Y^I = f^I (a), \quad \delta a = 0, \label{fifteen}
\end{equation}
where $f^I$ are arbitrary infinitesimal functions of one variable.  This is the
covariant version of the gauge symmetry of ${\cal L}_N$ that was used to argue that the
undesired solution of the equations of motion is pure gauge.

\medskip

\noindent{\it Reparametrization Invariant Action}

The formulas described above are not the whole story of the bosonic degrees of
freedom of the toroidally compactified heterotic string, because they lack the
Virasoro constraint conditions.  The standard way to remedy this situation is
to include an auxiliary world-sheet metric field 
$g_{\alpha\beta}(\xi)$, so that the world-sheet
Lorentz invariance is replaced by world-sheet general coordinate invariance.
Since we now want to include the coordinates $X^m$ describing the uncompactified
dimensions, as well, let us also introduce an induced world-sheet metric
\begin{equation}
G_{\alpha\beta} = g_{mn}(X) \partial_{\alpha}X^m\partial_{\beta} X^n,
\end{equation}
where $g_{mn}(X)$ is the string frame target-space metric in $d$ dimensions.
It is related to the canonically normalized metric by a factor of the form
${\rm exp} (\alpha\phi)$, where $\phi$ is the dilaton and $\alpha$ is a numerical
constant, which can be computed by requiring that the target-space lagrangian is
proportional ${\rm exp} (-2\phi)$.  We will mostly be
interested in taking $\phi$ to be a constant and $g_{mn}$ to be proportional
to the flat Minkowski metric. Then the heterotic string coupling constant
is $\lambda_H = {\rm exp}\, \phi$, and the desired world sheet lagrangian is
\begin{equation}
{\cal L}_g = - {1\over 2} \sqrt{-g} g^{\alpha\beta} G_{\alpha\beta}
+ {\tilde{Y} M \tilde{Y}\over 2\sqrt{-g} (\partial a)^2} - {\tilde{Y} L\,
\partial Y \cdot \partial a\over 2(\partial a)^2} . \label{Laux}
\end{equation}
Now, of course, $(\partial a)^2 = g^{\alpha\beta} \partial_\alpha a
\partial_\beta a$ and $ \partial Y \cdot \partial a = g^{\alpha\beta}
\partial_\alpha Y \partial_\beta a$.  The placement of the $\sqrt{-g}$ factors
reflects the fact that $\tilde{Y}/\sqrt{-g}$ transforms as a scalar. 

There are a few points to be made about ${\cal L}_g$.  First of all, the PST gauge
symmetries continue to hold, so it describes the correct degrees of freedom.
Second, just as for more conventional string actions, it has Weyl invariance:
$g_{\alpha\beta} \rightarrow \lambda g_{\alpha\beta}$ is a local symmetry.
This ensures that the stress tensor
\begin{equation}
T_{\alpha\beta} = - {2\over\sqrt{-g}} {\delta S_g\over\delta g^{\alpha\beta}},
\end{equation}
is traceless $(g^{\alpha\beta} T_{\alpha\beta} = 0)$.
Using the general coordinate invariance to choose $g_{\alpha\beta}$ conformally
flat, and using the PST gauge invariance to set $a = \xi^1$, the $Y$ equations
of motion reduce to those described in the previous subsection.  In addition,
one obtains the classical Virasoro constraints $T_{++} = T_{--} = 0$.  

The lagrangian ${\cal L}_g$ is written with an auxiliary world-volume metric, which is
called the Howe--Tucker or Polyakov formulation.  This is the most convenient
description for many purposes.  However, for the purpose of comparing to
expressions derived from the M5-brane later in this paper, it will be useful to
also know the version of the lagrangian in which the auxiliary metric is
eliminated --- the Nambu--Goto formulation.
Note that ${\cal L}_g$ only involves the metric components in the combination
$\sqrt{-g} g^{\alpha\beta}$, which has two independent components.
It is a straightforward matter to solve their equations of motion and 
eliminate them from the action. This leaves
the final form for the bosonic part of the heterotic string in
$10 - n$ dimensions
\begin{equation}
{\cal L} = - \sqrt{-G}\sqrt{1 + {\tilde{Y} M \tilde{Y}\over G(\partial a)^2} +
\left({\tilde{Y} L \tilde{Y}\over 2G(\partial a)^2}\right)^2 }- {\tilde{Y}
L\partial Y \cdot \partial a\over 2(\partial a)^2}, \label{hetfinal}
\end{equation}
where $G = \det G_{\alpha\beta}$, and now
\begin{equation}
(\partial a)^2 = G^{\alpha\beta} \partial_\alpha a \partial_\beta a, \quad \partial Y
\cdot \partial a = G^{\alpha\beta} \partial_\alpha Y \partial_\beta a.
\end{equation}

\medskip
\section{Wrapping the M-Theory Five-Brane on K3}

Let us now consider double dimensional reduction of the M5-brane 
on K3.\footnote{See ref.~\cite{aspinwall} for a review of the mathematics of K3
and some of its appearances in string theory dualities.} 
This is supposed to give the heterotic string in seven dimensions~\cite{witten,harvey,townsend}.
Our starting point is the bosonic part of the M5-brane action~\cite{perry}
in the general coordinate invariant PST formulation. 
Since the other 11d fields are still assumed to vanish, 
$g_{MN}(X)$ must be Ricci flat. We will take it to be a product of a Ricci-flat
K3 and a flat 7d Minkowski space-time.

Since the M5-brane is taken to wrap the spatial K3, 
the diffeomorphism invariances of
the M5-brane action in these dimensions can be used to equate 
the four world-volume  coordinates  that describe the K3 
with the four target-space coordinates that describe the K3. In other words,
we set $\sigma^{\mu} = ( \xi^{\alpha}, x^i)$ and $X^M = (X^m, x^i)$.
Note that Latin indices $i,j,k$ are used for the K3 dimensions $(x^i)$ 
and early Greek letters for the directions $(\xi^\alpha)$,
which are the world-sheet coordinates of the resulting string action.
This wrapping by identification of coordinates, together with the extraction of the
K3 zero modes, is what is meant by double dimensional reduction. 
With these choices, the 6d metric can be decomposed into blocks
\begin{equation}
\label{metricdecomposition}
\left(G_{\mu\nu}\right)=\left( \begin{array}{cc} 
                                    \tilde G_{\alpha\beta} & 0 \\ 
                                      0   &  h_{ij}    
                                          \end{array} \right) ,
\end{equation}
with $h_{ij}$ and $\tilde G_{\alpha\beta}$ being the K3 metric and the induced
metric on the string world-sheet, respectively. The purpose of the tilde is to
emphasize that $\tilde G_{\alpha\beta} = \tilde g_{mn} \partial_{\alpha} X^m
\partial_{\beta} X^n$, where $\tilde g_{mn}$ is the 7d part of the
canonical 11d metric. It differs from the metric introduced earlier by a scale factor,
which will be determined below. It is
convenient to take the PST scalar field $a$
to depend on the $\xi^{\alpha}$ coordinates only. This amounts to partially
fixing a gauge choice for the PST gauge invariance.

The two-form field $B$ has the following contributions from K3 zero modes:
\begin{equation}
\label{2formdecomposition}
B_{ij}= \sum_{I=1}^{22} Y^I(\xi) b_{Iij}(x), \;\; B_{\alpha i}=0, 
\;\; B_{\alpha\beta}=c_{\alpha\beta}(\xi),
\end{equation}
where $b_{Iij}$ are the 22 harmonic representatives of H${}^2$(K3, {\bf Z}), the
integral second cohomology classes of K3. Any other terms are either massive or can be
removed by gauge transformations.
The nonzero components of $H_{\mu\nu\rho}$ and $\Htilde^{\mu\nu}$ are
\begin{equation}
H_{\alpha i j}=\sum_{I=1}^{22} \partial_\alpha Y^I b_{Iij} \label{Hform1}
\end{equation}
\begin{equation}
\Htilde^{i j}=\sum_{I=1}^{22}\Ytilde^I \frac{1}{2}\ep^{i j k l}b_{Ik l}=
\sum_{I=1}^{22}\sqrt{h} \Ytilde^I (\ast b_I)^{i j}, \label{Hform2}
\end{equation}
where $\Ytilde^I=\ep^{\alpha\beta}\partial_\alpha Y^I \partial_\beta a$ as in 
eq.~(\ref{Ytilde}). Note that $c_{\alpha\beta}$ does not contribute.

Now we can compute the string action that arises from double dimensional
reduction by substituting the decompositions (\ref{Hform1}) and (\ref{Hform2})
into the five-brane Lagrangian. 
To make the connection with the heterotic string action of the previous section,
we make the identifications 
\begin{equation}
 L_{IJ}=\int_{K3} b_I\wedge b_J,
\end{equation}
\begin{equation}
M_{IJ}=\int_{K3} b_I\wedge \ast b_J.
\end{equation}
Note that $\ast b_I = b_J (L^{-1} M)^J{}_I$, and therefore $(L^{-1}M)^2 =1$,
as in sect. 2.
Note also that $b_I\wedge b_J$ and $b_I\wedge \ast b_J$ are closed four-forms, and
therefore they are cohomologous to the unique harmonic four-form of the K3, 
which is the volume form $\omega$. It follows that
\begin{equation} \label{bwedgeb}
b_I\wedge b_J=\ast b_I\wedge \ast b_J=\frac{L_{IJ}}{\cal V}\omega +dT_{IJ},\quad  
b_I\wedge \ast b_J=\frac{M_{IJ}}{\cal V}\omega +dU_{IJ},
\end{equation}
where ${\cal V}= \int_{K3} \omega$ is the volume of the K3
and $U_{IJ} =  T_{IK}(L^{-1} M)^K{}_J$. The exact terms are absent when either
two-form is self-dual, but there is no apparent reason why they should vanish when
both of them are anti-self-dual.
If we nevertheless ignore the exact pieces in these formulas,
substitute into  the Lagrangian,
and integrate over the K3, we obtain
\begin{equation}
\label{5brane on K3}
{\cal L}_1 = -{\cal V} \sqrt{-\tilde G}\sqrt{1+
\frac{\Ytilde^I M_{IJ} \Ytilde^J}{\tilde G(\partial a)^2 {\cal V}}
+\frac{1}{4}\left(\frac{\Ytilde^I L_{IJ}\Ytilde^J}
{\tilde G(\partial a)^2{\cal V}} \right)^2}
-\frac{\Ytilde^I L_{IJ} \partial_\alpha Y^J \tilde G^{\alpha\beta}\partial_\beta a}
{2(\partial a)^2}.
\end{equation}
This is precisely the heterotic string lagrangian (for $n=3$) presented in 
eq.~(\ref{hetfinal}) of the previous section provided that the 7d metric $g_{mn}$
in the string frame is related to the metric $\tilde g_{mn}$ derived from 11d by 
\begin{equation}
g_{mn} = {\cal V} \tilde g_{mn}
\end{equation}
so that $G_{\alpha\beta} = {\cal V}
\tilde G_{\alpha\beta}$. This is the same scaling rule found by a different 
argument in ref.~\cite{witten}.  Then, following ref.~\cite{witten},
the Einstein term in the 7d lagrangian
is proportional to ${\cal V} \sqrt{- \tilde g} R(\tilde g) = {\cal V}^{-3/2}
\sqrt{- g} R(g)$, from which
one infers that ${\cal V} \sim \lambda_H^{4/3}$.

To complete the argument we must still explain why terms that have been dropped
make negligible contributions. It is not at all obvious that 
the exact pieces in eq.~(\ref{bwedgeb}) can be neglected, but it is what is required
to obtain the desired answer. The other class of terms that have been dropped are
the Kaluza--Klein excitations of the five-brane on the K3. By simple dimensional
analysis, one can show that in the heterotic string metric these contributions 
to the mass-squared of excitations are of order $\lambda_H^{-2}$. Therefore they
represent non-perturbative corrections from the heterotic viewpoint. Since our
purpose is only to reproduce the perturbative heterotic theory, they can be dropped.
Another class of contributions, which should not be dropped, correspond to
simultaneously wrapping the M2-brane around a 2-cycle of the K3. These wrappings
introduce charges for the 22 U(1)'s, according to how many times each cycle is wrapped.
The contribution to the mass-squared of excitations depends on the shape of the K3,
of course, but in the heterotic metric it is independent of its volume and hence of the
heterotic string coupling constant.

\noindent{\it Acknowledgment}

I am grateful to M. Aganagic, S. Cherkis, 
J. Park, M. Perry, and C. Popescu for collaborating on portions of this work.
This work is 
supported in part by the U.S. Dept. of Energy under Grant No.
DE-FG03-92-ER40701.

\end{document}